\documentclass[aps,pre,reprint]{revtex4-1}
\usepackage[T1]{fontenc}
\usepackage{amsmath}
\usepackage[latin1]{inputenc}
\usepackage{graphicx}

\usepackage{epsfig}

\usepackage{color}

\begin{document}
\title{Bicontinuous and mixed gels in binary mixtures of patchy colloidal particles}

\author{Daniel de las Heras}
\email{delasheras.daniel@gmail.com}
\affiliation{Centro de F\'{i}sica Te\'orica e Computacional da Universidade de Lisboa, Avenida Professor Gama Pinto 2, P-1649-003, Lisbon, Portugal}

\author{Jos\'e Maria Tavares}
\email{josemaria.castrotavares@gmail.com}
\affiliation{Instituto Superior de Engenharia de Lisboa, Rua Conselheiro Em\'{\i}idio Navarro, P-1590-062 Lisbon, Portugal, and Centro de F\'{i}sica Te\'orica e Computacional da Universidade de Lisboa, Avenida Professor Gama Pinto 2, P-1649-003, Lisbon, Portugal}

\author{Margarida M. Telo da Gama}
\email{margarid@cii.fc.ul.pt}
\affiliation{Departamento de F\'{\i}sica, Faculdade de Ci\^encias da Universidade de Lisboa, Campo Grande, P-1749-016, Lisbon, Portugal, and Centro de F\'{i}sica Te\'orica e Computacional da Universidade de Lisboa, Avenida Professor Gama Pinto 2, P-1649-003, Lisbon, Portugal}

\date{\today}
\begin{abstract}
We investigate the thermodynamics and percolation regimes of model binary mixtures of patchy colloidal particles. The particles of each species have three sites of 
two types, one of which promotes bonding of particles of the same species while the other promotes bonding of different species. We find up to four percolated 
structures at low temperatures and densities: two gels where only one species percolates, a mixed gel where particles of both species percolate but neither species 
percolates separately, and a bicontinuous gel where particles of both species percolate separately forming two interconnected networks. The competition between the 
entropy and the energy of bonding drives the stability of the different percolating structures. Appropriate mixtures exhibit one or more connectivity transitions 
between the mixed and bicontinuous gels, as the temperature and/or the composition changes. 
\end{abstract}

\maketitle

\section{Introduction \label{introduction}}
Colloidal self-assembly provides a route for the fabrication of functional materials with novel mesoscopic structures, which are robust but adaptable to changing external conditions. Indeed, soft colloidal solids including gels and glasses, are examples of three dimensional structures that may be tuned, e.g., by changes in temperature.    

Advances in the synthesis of well-defined colloidal particles, with surface patterning in the sub-micron range, paves the way for the design of their macroscopic behaviour  \cite{vanblaaderen2006,intro1,Granick1,Granick2,Sacanna1,Sacanna2,B718131K,Leunissen1}. 
One focus of research concerns the development of large-scale fabrication techniques required for the exploitation of the new materials in a wide range of applications \cite{intro2}. The other, which includes modeling and computational efforts, based on the fact that anisotropic colloids may be viewed as the molecules of novel materials, aims at  tailoring their (self)assembly into a variety of functional structures \cite{Glotzer15102004,intro3}.

The primitive model of patchy colloids consists of hard-spheres with $f$ patches on their surfaces. The patches act as bonding sites and promote the formation of well defined clusters, whose structure and size distribution depend on the properties of the patches ($f$ and the bonding energy) and on the thermodynamic conditions (density and temperature). Computational studies of patchy particle models, over the last decade, produced a number of results \cite{SciortinoReview2010}. In particular, Sciortino and co-workers showed that, for low values of $f$ (approaching 2), low densities and temperatures can be reached without encountering the phase boundary. These empty phases were shown to be network (percolated) liquids, suggesting that, on cooling, patchy particles with low functionality assemble into amorphous states of arbitrary low density -- equilibrium gels. These gels are novel structures that may be assembled under controlled and reproducible equilibrium conditions. 

Remarkably, the results of the simulations are well described by classical liquid state theories: Wertheim's first order perturbation theory \cite{wertheim1,wertheim2,wertheim3,wertheim4} predicts correctly the equilibrium thermodynamic properties; Flory-Stockmayer \cite{flory1,stock1,flory2} theories of polymerization describe quantitatively the size distributions of the patchy particle clusters as well as the percolation threshold.

Very recently, Ruzicka and co-workers reported experimental evidence of empty liquids in dilute suspensions of Laponite \cite{emptyexp} confirming the patchy particle model as the primitive model of real equilibrium gels.
In this paper we use theoretical tools to study the structure of  equilibrium gels in model binary mixtures of patchy colloids. We investigate the stability of these amorphous soft-solid materials and show that there are four types of gels, the stability of which may be controlled by the temperature and composition of the mixture.

The phase diagram of mixtures of patchy particles with identical patches was investigated by  Bianchi et al \cite{PhysRevLett.97.168301} and established the stability of the empty fluid regime. Later on we clarified the conditions for the stability of empty fluids and characterized their (network) structure using a generalization of the Flory-Stockmayer theory  \cite{C0SM01493A}. The phase diagram of monodisperse systems and of mixtures of colloidal particles with dissimilar patches was also investigated, elucidating the conditions for the emergence of criticality and establishing more general conditions for the stability of network (percolated) fluids \cite{C0SM01493A,heras:104904,PhysRevLett.106.085703,10.1063/1.3605703}.

In the context of a related model, Hall and co-workers \cite{B907873H,goyal:064511} investigated the structures formed by monodisperse systems and equimolar mixtures of dipolar colloids with particles of two different sizes and dipole moments. Their studies, based on (discontinuous) molecular dynamics, focused on kinetic pathways and the resulting non-equilibrium structures. In addition to solid phases, they report two novel structures, both bicontinuous gels. The bicontinuous gels were shown to have tunable pore size determined by the size and dipole moment ratios of the colloidal particles.
 


Here we investigate the thermodynamic phase diagram and the percolation regime of a model binary mixture of patchy colloidal particles. Each species has three patches with two different types of bonding energies, one promoting bonds between the same species while the other promotes bonds between different species. We have identified four distinct percolated phases: two gels where only one of the species is percolated, a mixed gel where the two species are percolated but neither species percolates by itself, and a bicontinuous gel where the two species percolate independently, forming two interpenetrating spanning networks. The competition between the entropy of bonding and the energy of bonding determines the stability of the different network phases. We show that for appropriate mixtures a transition between the mixed and bicontinuous gels is induced as the temperature and/or the composition of the mixture changes.

The remainder of the paper is organised as follows. In section \ref{theory} we describe the model and the theory: Wertheim's thermodynamic perturbation theory for mixtures of associating fluids (\ref{wertheim}) and a generalization of the Flory-Stockmayer theory of percolation to mixtures of patchy particles (\ref{percolationsec}). In section \ref{Results} we present the results for a series of representative mixtures. We describe the fluid phase diagrams, with emphasis on the percolation regime, of symmetric and asymmetric mixtures and discuss the conditions for the connectivity transitions between mixed and bicontinuous gels. Finally, in Sec. \ref{Conclusions} we summarize our conclusions.

\section{Model and Theory\label{theory}}

The model is a binary mixture of $N_1$ and $N_2$ equisized hard spheres (HSs) with diameter $\sigma$. Each species has three interaction or bonding sites on its surface. The 
association between bonding sites is described using Wertheim's first-order perturbation theory. Therefore, two particles can form one single bond between two sites, one on 
each particle. Bonding sites are distributed randomly on the particle surfaces in such a way that all sites are available to bonding (i.e. there is no shading of any site by 
nearby bonds).

\begin{figure}
\epsfig{file=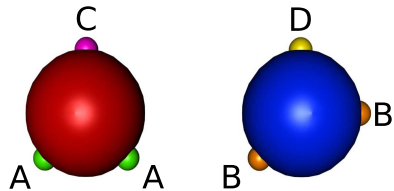,width=2.in,clip=}
\caption{The model: a binary mixtures of equisized hard spheres with three bonding sites of two different types on each particle.} 
\label{fig1}
\end{figure}

Species $1$ has two bonding sites of type $A$ and one of type $C$ while species $2$ has two sites of type $B$ and one of type $D$ (see Fig. \ref{fig1}). Although the model may 
seem complex, the idea is quite simple: each species has two types of bonding sites, one which promotes bonds between particles of the same species (sites of type $A$ and $B$) 
while the other promotes bonds between different species (sites of type $C$ and $D$). Note that the minimum number of bonds required for each species to percolate is $3$. 

\subsection{Helmholtz free energy: Wertheim's thermodynamic perturbation theory \label{wertheim}}

In references \cite{wertheim1,wertheim2,wertheim3,wertheim4,Chapman:1057} Wertheim's first order perturbation theory is described in detail. Here we state briefly the main 
results and set the notation. 

The Helmholtz free energy of the binary mixture of patchy particles can be written as the sum of two contributions:
\begin{eqnarray}
f_{H}=F/N=f_{HS}+f_b,
\end{eqnarray}
where $f_{HS}$ is the free energy density of the reference system of hard-spheres and $f_b$ the perturbation due to the attractive bonding interactions. $N=N_1+N_2$ is the total 
number of particles. 

The free energy of the reference system of HSs, $f_{HS}$, may be split into ideal-gas and excess terms: $f_{HS}=f_{id}+f_{ex}$. The ideal-gas free energy is given (exactly) by
\begin{equation}
\beta f_{id}=\ln\eta-1+\sum_{i=1,2}x^{(i)}\ln(x^{(i)} {\cal V}_i),
\end{equation}
where $\beta=kT$ is the inverse thermal energy, ${\cal V}_i$ is the thermal volume, $x^{(i)}=N_i/N$ is the molar fraction of species $i$ and $\eta=\eta_1+\eta_2$ is the total packing fraction ($\eta=v_s\rho$, with $\rho$ the total number density and $v_s=\pi/6\sigma^3$ the volume of a single particle). The excess part accounts for the excluded volume interactions between hard spheres. We have approximated it using the Carnahan-Starling equation of state for HSs \cite{carnahan:635} (note that both species have the same diameter):
\begin{equation}
\beta f_{ex}=\frac{4\eta-3\eta^2}{(1-\eta)^2.}
\end{equation}
 
The bonding free energy, $f_b$, accounts for the interactions between sites. Within Wertheim's first-order perturbation theory \cite{wertheim1,wertheim2,wertheim3,wertheim4}, it 
is written as \cite{Chapman:1057}:
\begin{equation}
\beta f_b=\sum_{i=1,2} x^{(i)}\left[\sum_{\alpha\in \Gamma(i)}\left(\ln X_\alpha^{(i)}-\frac{X_\alpha^{(i)}}{2}\right)+\frac12 n(\Gamma(i))\right],\label{fb}
\end{equation}
where  $\Gamma(i)$ is the set of bonding sites or patches on one particle of species $i$ ({\it i.e.}, $\Gamma(1)=\{A,A,C\}$, $\Gamma(2)=\{B,B,D\}$) and $n(\Gamma(i))$ is the total number of bonding sites per particle of species $i$. The variables $\{ X_\alpha^{(i)} \}$ are the probabilities of finding one site of type $\alpha$ on a particle of species $i$ {\it not} bonded. The bonding free energy has two contributions: the bonding energy and an entropic term related to the number of ways of bonding two particles. 

The law of mass action establishes a relation between $\{ X_\alpha^{(i)} \}$ and the thermodynamic variables: 
\begin{equation}
X_\alpha^{(i)}=\left[1+\eta\sum_{j=1,2}x^{(j)}\sum_{\gamma\in \Gamma(j)}X_\gamma^{(j)}\Delta_{\alpha\gamma}^{(ij)}\right]^{-1}.\label{xnotbonded}
\end{equation}
$\Delta_{\alpha\gamma}^{(ij)}$ characterises the bond between a site $\alpha$ on a particle of species $i$ and a site $\gamma$ on a particle of species $j$. We model the interaction between sites by square well potentials with depths $\varepsilon_{\alpha\gamma}$ that depend on the type of bonding sites ($\alpha$ and $\gamma$) but not on the particle species ($i$ and $j$). As a result, when the particles have the same diameter, $\Delta_{\alpha\gamma}^{(ij)}$ are independent of the particle species, and are given by
\begin{equation}
\Delta_{\alpha\gamma}^{(ij)}=\Delta_{\alpha\gamma}=\frac1{v_s}\int_{v_{b}}g_{HS}({\bf r})\left[\exp(\beta\varepsilon_{\alpha\gamma})-1\right]d{\bf r}.\label{delta}
\end{equation}
$g_{HS}({\bf r})$ is the radial distribution function of the reference HS fluid and the integral is calculated over the bond volume $v_b$ (we have considered that all bonds have the same volume, $v_b=0.000332285\sigma^3$). We approximate $g_{HS}$ by its contact contact value (the bond volume is very small). Then Eq. (6) simplifies to:
\begin{equation}
\Delta_{\alpha\gamma}=\frac{v_b}{v_s}\left[\exp(\beta\varepsilon_{\alpha\gamma})-1\right]A_0(\eta),\label{delta2}
\end{equation}
where
\begin{equation}
A_0(\eta)=\frac{1-\eta/2}{(1-\eta)^3}
\end{equation}
is the contact value of $g_{HS}$. By substituting $\Delta_{\alpha\gamma}$ given by Eq. (\ref{delta2}) in Eq. (\ref{xnotbonded}) we find that $X_\alpha^{(i)}$ depends only on $\alpha$, the type of site ({\it i.e.}, $X_\alpha^{(i)}=X_\alpha,\;\; \forall\; i$).

In what follows we will denote the composition of the mixture $x$ by the molar fraction of species $1$: $x\equiv x^{(1)}$ ($x^{(2)}=1-x$). We choose to minimize the Gibbs free energy per particle ($g=p/\rho+f_H$) to obtain the equilibrium properties of the mixture. We set the composition $x$, pressure $p$ and temperature $T$ and locate the binodals by a standard common-tangent construction on $g(x)$.

\subsection{Percolation thresholds\label{percolationsec}}

\begin{figure}
\epsfig{file=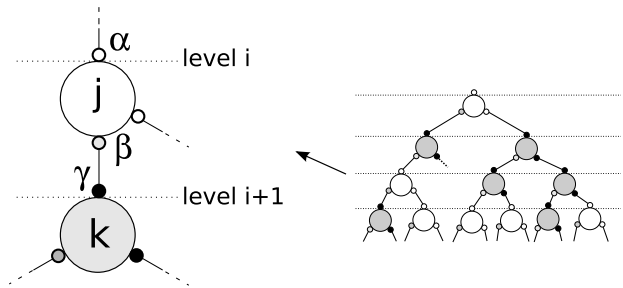,width=3.2in,clip=}
\caption{Schematic representation of a tree tree-like cluster of a binary mixture of patchy particles (right). On the left one particle of species $k$ (at level $i+1$) is bonded to a particle of species $j$ (at level $i$) through sites of type $\gamma$ and $\beta$ respectively.} 
\label{treelike}
\end{figure}

The percolation threshold is analysed using a generalization of the Flory-Stockmayer theory of polymerization \cite{flory1,stock1,flory2} (see Refs. \cite{PhysRevE.81.010501,C0SM01493A} for details) which neglects closed loops. The percolation thresholds and the cluster size distributions were found to be in good agreement with the results of Monte Carlo simulations for the pure fluids \cite{tavares3}, and we expect the same level of accuracy for binary mixtures. 

Let $n_{i+1,\gamma}^{(k)}$ denote the number of bonded sites of type $\gamma$ on particles of species $k$ at the level $i+1$ of a tree-like cluster of particles (see Fig. \ref{treelike}). This is related to $\{ n_{i,\alpha}^{(j)} \}$, the set of all types of bonded sites in the previous level, through the recursive relations given by (see Fig. \ref{treelike} and Refs. \cite{PhysRevE.81.010501,C0SM01493A}):
\begin{equation}
n_{i+1,\gamma}^{(k)}=\sum_{j}\sum_{\alpha\in\Gamma_d(j)}\sum_{\beta\in\Gamma_d(j)}p_{\beta_j\rightarrow\gamma_k}\left(f_{\beta}^{(j)}-\delta_{\beta\alpha}\right)n_{i,\alpha}^{(j)},
\label{levelperco}
\end{equation}
where: $\Gamma_d(j)$ is the the set of distinct bonding sites on species $j$ ({\it i.e.}, $\Gamma_d(1)=\{A,C\}$ and $\Gamma_d(2)=\{B,D\}$); the sum on $j$ runs over all species $j=\{1,2\}$; $f_{\beta}^{(j)}$ is the number of $\beta$ sites on a particle of species $j$ ({\it e.g.}, $f_A^{(1)}=2$); $\delta_{\beta\alpha}$ is the Kronecker delta and $p_{\beta_j\rightarrow\gamma_k}$ is the probability of bonding a site $\beta$ on a particle of species $j$ to a site $\gamma$ on a particle of species $k$. Summing over all $k$ and $\gamma$, we obtain
\begin{equation}
P_{\beta_j}=\sum_k\sum_{\gamma\in\Gamma_d(k)}p_{\beta_j\rightarrow\gamma_k},
\end{equation}
the probability of finding a bonded site $\beta$ on a particle of species $j$. At this point, it is possible to establish a connection with thermodynamics through the law of mass action, since by definition: 
\begin{equation}
P_{\beta_j}=1-X_\beta^{(j)}.
\label{eqptotal}
\end{equation}
A term-by-term analysis of Eqs. (\ref{eqptotal}) and (\ref{xnotbonded}) yields expressions for the equilibrium probabilities $p_{\beta_j\rightarrow\gamma_k}$ in terms of the thermodynamic variables. 

In order to calculate the percolation threshold it is convenient to write equations (\ref{levelperco}) in matrix form:
\begin{equation}
\tilde n_i=\tilde T^i\tilde n_0,\label{progressions}
\end{equation}
where $\tilde n_i$ is a vector with components $n_{i,\gamma}^{(k)}$ and $\tilde T$ is a square matrix of size $\Gamma_d(1)+\Gamma_d(2)$ with:
\begin{equation}
T_{\gamma_k\alpha_j}=\sum_{\beta\in\Gamma_d(j)}p_{\beta_j\rightarrow\gamma_k}\left(f_{\beta}^{(j)}-\delta_{\beta\alpha}\right).
\end{equation}

For the binary mixture of particles with three sites, $\tilde T$ is a $4\times4$ matrix \footnote{In the model considered here the subscripts denoting the particle species in $p_{\beta_j\rightarrow\gamma_k}$ are redundant, and have been omitted in what follows. $p_\beta\gamma$ is the probability of bonding a site $\beta$ to a site $\gamma$.}:
\begin{eqnarray}
\tilde T  & = & \left(\begin{array}{c | c} \tilde T_{11} & \tilde T_{21} \\ \hline \tilde T_{12} & \tilde T_{22}  \end{array} \right) =  \nonumber \\ 
          & = &  \left(\begin{array}{c c | c c} p_{AA} + p_{CA} & 2p_{AA} & p_{BA} + p_{DA} & 2p_{BA}  \\ p_{AC} + p_{CC} & 2p_{AC} & p_{BC} + p_{DC} & 2p_{BC}  \\ \hline p_{AB} + p_{CB} & 2p_{AB} & p_{BB} + p_{DB} & 2p_{BB}  \\ p_{AD} + p_{CD} & 2p_{AD} & p_{BD} + p_{DD} & 2p_{BD}   \end{array}\right). 
\label{matrix}
\end{eqnarray}

\begin{figure*}
\epsfig{file=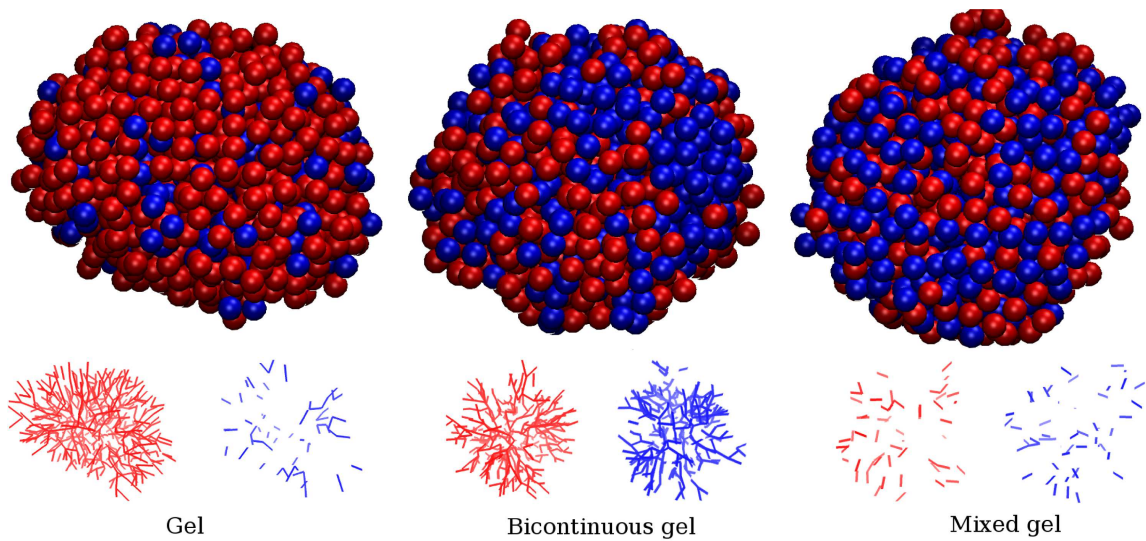,width=6.in,clip=}
\caption{Schematic representation of the gel structures: a standard gel of species $1$ (left), a bicontinuous gel or bigel (middle), and a mixed gel (right). The figure is 
a cartoon consistent with the bonding rules of the model. Each particle is bonded to three other particles (bonding sites not shown) and there are no closed loops. The upper 
row depicts the position of the particles. Bonds between particles of the same species are represented in the lower row by straight lines connecting the center of mass of two 
bonded particles. Each red (blue) line corresponds to a bond between two particles of species $1$ $(2)$. Bonds between particles of different species are not shown.}
\label{figschematic}
\end{figure*}

Let $\lambda_+$ denote the largest (absolute value) eigenvalue of $\tilde T$. Then, the progressions defined by Eq. (\ref{progressions}) converge to $0$ if $|\lambda_+|<1$ and 
diverge if $|\lambda_+|>1$. The percolation threshold occurs when $|\lambda_+|=1$. The analysis of $\lambda_+$ allows us to distinguish percolated from non-percolated structures of 
the mixture, but it does not differentiate, for example, bicontinuous from mixed gels as the mixture is percolated in both cases. In order to distinguish different percolated 
structures we have to analyse, in addition to the percolation threshold of the mixture, the percolation thresholds of each species. The matrix $\tilde T$ may be split into four 
blocks $\tilde T_{ij}$, $i,j=\{1,2\}$ (see Eq. (\ref{matrix})), where each block $\tilde T_{ij}$ is a $2\times2$ matrix that accounts for (all) the bonds between species $i$ and 
$j$. The largest eigenvalues of $\tilde T_{11}$ and $\tilde T_{22}$ ($\lambda_+^{(11)}$ and $\lambda_+^{(22)}$ respectively) yield the percolation thresholds of both species. In 
a percolated binary mixture ($|\lambda_+|\ge1$) it is then possible to distinguish four different percolating structures:

\begin{itemize}
\item $|\lambda_+^{(11)}<1|$ and $|\lambda_+^{(22)}<1|$. The mixture is percolated but removal of one of the species destroys the connectivity of the spanning cluster. 
We refer to this structure as a mixed gel ($MG$) and illustrate it schematically in Fig. \ref{figschematic} (right).
\item $|\lambda_+^{(11)}\ge1|$ and $|\lambda_+^{(22)}\ge1|$. The mixture as well as both species are percolated. We refer to this structure as a bicontinuous gel or ''bigel'' 
($BG$) as the spanning cluster consists of two interconnected spanning clusters of the pure components. Removal of one species does not destroy the connectivity of the 
spanning cluster. The bigel is illustrated in the middle of Fig. \ref{figschematic}.
\item $|\lambda_+^{(11)}\ge1|$ and $|\lambda_+^{(22)}<1|$. The mixture and species $1$ are percolated but species $2$ is not. Removal of species $2$ does not destroy the 
connectivity of the spanning cluster. We refer to this structure as a standard gel $G_1$ 
and illustrate it schematically in Fig. \ref{figschematic} (left).
\item $|\lambda_+^{(11)}<1|$ and $|\lambda_+^{(22)}\ge1|$. The mixture and species $2$ are percolated but species $1$ is not. Removal of species $1$ does not destroy the 
connectivity of the spanning cluster. We refer to this structure as a standard gel $G_2$. 
\end{itemize} 

Table \ref{table1} classifies the structure of the binary mixture based on the analysis of the percolating cluster.

\begin{table}[h]
\begin{center}
\vspace{.1in}
\begin{tabular}{|c|c|c|c|}
\hline
$|\lambda_+|$ & $|\lambda_+^{(11)}|$ & $|\lambda_+^{(22)}|$ & percolation state \\
\hline
$<1$ & $<1$ & $<1$ & non-percolated \\
$\ge1$ & $<1$ & $<1$ & percolated: mixed gel \\
$\ge1$ & $\ge1$ & $\ge1$ & percolated: bigel \\
$\ge1$ & $\ge1$ & $<1$ & percolated: gel 1 \\
$\ge1$ & $<1$ & $\ge1$ & percolated: gel 2 \\
\hline
\end{tabular}
\caption{Structure of the binary mixture based on the type of percolating cluster, characterised by the largest eigenvalues of $\tilde T$, $\tilde T_{11}$ 
and $\tilde T_{22}$.}\label{table1}
\end{center}
\end{table}

\section{Results}\label{Results}

The phase diagram of the pure fluids was analysed in Ref. \cite{C0SM01493A}. The fluids exhibit a first order phase transition that ends at a critical point. The transition 
involves two fluid phases at different densities and fraction of unbonded sites. One phase (high density, small fraction of unbonded sites) is always percolated while the other 
(low density and large fraction of unbounded sites) is percolated only at pressures and densities close to the critical point. We call these phases network fluid and vapor, 
respectively.

The results for the binary mixture are presented as temperature-composition phase diagrams at constant pressure and the following remarks apply to all of them: the 
binodal lines are depicted as solid-black lines; shaded areas are two-phase regions; empty circles are critical points; black squares are azeotropic points. In all cases the 
network fluid phases are percolated (the liquid side of the binodal is percolated). Therefore, at a given composition, the mixture is percolated at temperatures below the temperature 
of the network fluid-vapor phase transition. In order to distinguish different percolated structures, percolation thresholds for each species are also calculated. The percolation line of species $1$ is depicted as a solid-red line. When we remove the bonds between species $1$ and $2$, species $1$ is percolated on the right of this line (indicated by the red arrow). The dashed-blue line is the percolation line of species $2$. On the left of this line (indicated by the blue arrow), species $2$ is percolated when we remove the bonds between species $2$ and $1$.

\subsection{Symmetric binary mixture AAA-BBB}\label{AAABBB}
We start by describing the results for a simplified model: a binary mixture where the particles of species $1$ have three patches of type $A$ and those of species $2$ have three 
patches of type $B$. In addition, we analyse the symmetric case only. That is, we set $\varepsilon_{AA}=\varepsilon_{BB}=\varepsilon$ and vary the bonding energy of the $AB$ bonds 
($\varepsilon^*_{AB}=\varepsilon_{AB}/\varepsilon_{AA}$). 

\begin{figure}
\epsfig{file=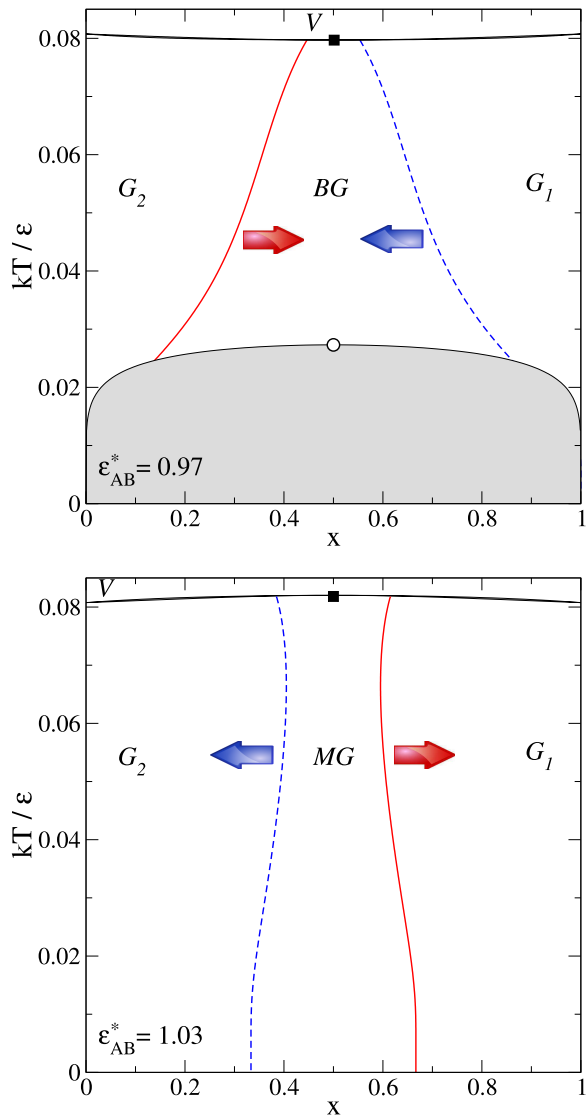,width=3.in,clip=}
\caption{Reduced temperature-composition phase diagram at constant pressure for a symmetric $AAA-BBB$ binary mixture with $\varepsilon_{AB}^*=0.97$ (top) and $\varepsilon_{AB}^*=1.03$ (bottom). The pressure is $p^*=pv_s/\varepsilon_{AA}=4.19\times10^{-5}$, well below the critical pressure of the pure fluids. $x$ is the composition of species $1$ (HSs with three patches of type $A$). See the beginning of section \ref{Results} for a description of the symbols and the graphical codes.} 
\label{fig4}
\end{figure}

Two representative phase diagrams of this mixture are depicted in Fig. \ref{fig4} for different bonding energies: $\varepsilon^*_{AB}=0.97$ (top) and $\varepsilon^*_{AB}=1.03$ (bottom). 

In the first mixture (top panel) the interaction energy between sites on different species is lower than the interaction energy between sites on the same species ($\varepsilon^*_{AB}=0.97$). At high temperatures there is a network fluid-vapor phase transition with a positive azeotrope at $x=0.5$. At low temperatures a demixing region, bounded by an upper critical point, occurs between two network fluids. The demixing is driven by the bonding energy: at low temperatures most sites are bonded but $AB$ bonds hardly occur as they increase the bonding energy. Obviously, the same effect occurs at intermediate temperatures (below the network fluid-vapor phase transition and above the demixing region). However, in this range of temperatures the level of association is much lower and the energy gain does not compensate the loss in the entropy of mixing. As a result, the mixture is stable. 

Let us focus now on the percolation threshold(s). Below the network fluid-vapor phase transition the mixture is percolated. The analysis of the percolation threshold for each species (solid-red and dashed-blue lines) reveals three different gel structures. In the region located to the left of the solid-red line ($x\gtrsim0$) only the particles of species $2$ are percolated. The gel is a standard ($G_2$). The same behaviour occurs (recall that the mixture is symmetric) to the right of the shaded-blue line ($x\lesssim1$), but in this case the 
gel is a standard ($G_1$). In the intermediate region, both species percolate separately, there is a bicontinuous gel ($BG$). As the temperature decreases, the BG region spans a wider range of compositions. At very low temperatures, however, the bigel is pre-empted by the demixing region where two standard gels $G_1$ and $G_2$ coexist. 

The second mixture, Fig. \ref{fig4} (bottom panel), illustrates the behaviour of a compound forming mixture: the attraction between sites on different species is stronger than that between sites on the same species. This has a profound effect both on the thermodynamic and the percolation behaviour. The network fluid-vapor phase transition and the azeotropic point at $x=0.5$ are still present. However, this is a negative azeotrope, characteristic of compound forming mixtures. Another important difference is that the mixture is always stable at temperatures below the network fluid-vapor phase transition. The demixing region disappears as there is no driving force for phase separation. The bonding free energy (entropy 
of mixing) is minimal (maximal) when the number of $AB$ bonds is maximal. 

At the percolation level there are also important differences. In this mixture, there is no bigel structure. Near $x=0.5$ neither species percolates separately although the mixture 
is percolated. This is the mixed gel ($MG$) structure where the $AB$ bonds dominate being responsible for the connectivity of the spanning cluster. 
As the composition of the mixture increases, the fraction of bonds between particles of species $1$ ($AA$ bonds) also increases with a corresponding decrease in the fraction of $AB$ bonds. As a result the $MG$ structure is replaced by a standard $G_1$ gel. The symmetric behaviour (formation of a $G_2$ gel) occurs at low values of the composition. 

The percolation thresholds of species $1$ and $2$ tend asymptotically to $x=2/3$ and $x=1/3$ respectively, as the temperature vanishes. For mixtures with compositions $x<1/3$ or $x>2/3$ 
there are no $MG$ structures, as the fraction of $AB$ bonds is small and the gel is a standard single species structure. This is not the case for bicontinuous gels. At low 
temperatures the percolation thresholds of species $1$ and $2$ tend asymptotically to $x=1$ and $x=0$ (note that in Fig. \ref{fig4} (top) the percolation lines are not shown at 
temperatures below the demixing region). If there was no demixing, a bicontinuous gel would be found at any composition at low temperatures.   

The symmetric mixture analysed above is the simplest system where bicontinuous and mixed gels are formed. However, it is not possible to find both structures in the same mixture. In addition, the bicontinuous structure is pre-empted by a demixing transition between two standard gels. To address these questions, we analyse a (more) general class of $AAC-BBD$ mixtures in the next section.

\subsection{Symmetric binary mixture AAC-BBD}\label{AACBBD}

\begin{figure}
\epsfig{file=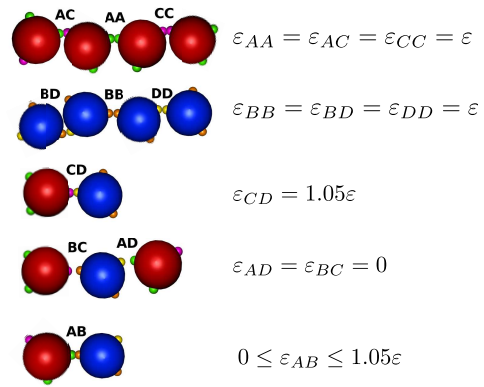,width=2.5in,clip=}
\caption{Bonding energies between different sites of the model.} 
\label{figinteractions}
\end{figure}

There are up to $10$ bonding energies in $AAC-BBD$ binary mixtures and, therefore, many possibilities to describe the formation of bicontinuous and mixed gels. We have focused on 
the simplest by restricting the bonding energies as depicted in Fig. \ref{figinteractions}: The bonding interaction between any pair of sites on particles of the same species 
is identical ($\varepsilon_{AA}=\varepsilon_{AC}=\varepsilon_{CC}=\varepsilon_{BB}=\varepsilon_{BD}=\varepsilon_{DD}=\varepsilon$). This sets the energy scale of the model; The strongest interaction is that between sites of type $C$ and $D$ ($\varepsilon_{CD}=1.05\varepsilon$); There is no interaction between sites of type $A$ and $D$ nor between sites $B$ 
and $C$ ($\varepsilon_{AD}=\varepsilon_{BC}=0$); Finally, we vary $\varepsilon_{AB}$, the interaction between sites of type $A$ and $B$. The physical idea behind this choice is the following: if the interaction between unlike species is due to the strongest $CD$ bonds only ($\varepsilon_{AB}=0$), a bicontinuous gel structure is expected at low temperatures. 
The reason is that most of the particles will be bonded to two other particles of the same species and to one particle of the other species, resulting in long interconnected chains 
of identical particles. On the other hand, if the interaction between $A$ and $B$ sites is sufficiently strong a mixed gel is expected as many particles are bonded to at least two particles of the other species (through $CD$ and $AB$ bonds).

In what follows we describe temperature-composition phase diagrams for different values of the interaction between the $A$ and $B$ sites. The pressure is  $p^*=pv_s/\varepsilon=4.19\times10^{-5}$, the same as in the previous section ({\it i.e.} well below the critical pressure of the pure fluids).

\subsubsection{$\varepsilon_{AB}\ll\varepsilon$}

\begin{figure}
\epsfig{file=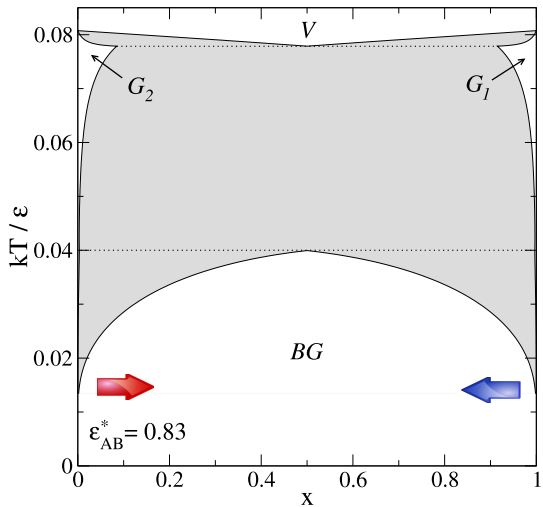,width=3.in,clip=}
\caption{Reduced temperature-composition phase diagram at constant pressure for an $AAC-BBD$ binary mixture with $\varepsilon_{AB}^*=0.83$. The pressure is $p^*=4.19\times10^{-5}$, 
well below the critical pressure of the pure fluids. $x$ is the molar fraction for species $1$.} 
\label{fig0.83}
\end{figure}

In Fig. \ref{fig0.83} we illustrate the phase diagram for a binary mixture with $\varepsilon_{AB}^*=\varepsilon_{AB}/\varepsilon=0.83$. The behaviour is qualitatively the same for 
lower values of the $AB$ interaction down to $\varepsilon_{AB}^*=0$. At high temperatures and large (small) values of the composition there is a phase transition between a vapor 
and a gel phase where only species $1$ ($2$) is percolated. The regions of stability of $G_1$ and $G_2$ gels are relatively small due to the presence of a large demixing region 
which meets the fluid-vapor binodal at a $G_1VG_2$ triple point. As the temperature decreases, the two-phase region increases until it spans a very large range of compositions. 
However, at sufficiently low temperatures, a new region of stability appears. Based on the percolation analysis (the percolation lines are not shown as they are always inside the demixing region) this new low temperature phase is a bicontinuous gel. The phase diagram exhibits a first order phase transition between $G_1$ or $G_2$ and the $BG$ phase.  

The phase behaviour can be rationalized in terms of the competition between the entropy of bonding and the energy of bonding. At very low temperatures, the energy of bonding dominates. The mixture is stable due to the formation of $CD$ bonds which are energetically favourable. The stable phase is a bicontinuous gel since each particle is bonded to a single particle 
of the other species. The other two bonding sites (type $A$ for species $1$ or type $B$ for species $2$) are bonded to identical sites, on particles of the same species, giving rise to two interconnected spanning networks. The reason is that $AA$ and $BB$ bonds are stronger than $AB$ bonds. Strictly speaking, the previous argument is correct only at $x=0.5$ and very low temperatures. If $x>0.5$, for example, it is also possible to find bonds of type $CC$ or $AC$ since there are more sites of type $C$ than sites of type $D$. However, it is only at values of the composition very close to those of the pure fluids that the $BG$ phase is replaced by standard $G_1$ or $G_2$ gels. At high temperatures, the behaviour of the system is dominated by the entropy of bonding. $CD$ bonds are energetically favourable, but the formation of these bonds decreases the entropy of bonding since the other two bonding sites are forced to form bonds with identical sites ($AB$ bonds are still unfavourable). Note that the same argument applies at low temperatures, but at high temperatures the gain in the energy of bonding from $CD$ bonds does not compensate the loss in the entropy of bonding (in addition, the fraction of unbonded sites is higher at high temperatures). The entropy of bonding drives the phase separation between the two standard gels $G_1$ and $G_2$ where bonds between sites of identical particles are maximized, independently of their type ({\it e.g.} $AA$, $AC$ or $CC$ bonds for the $G_1$ phase).  

\begin{figure}
\epsfig{file=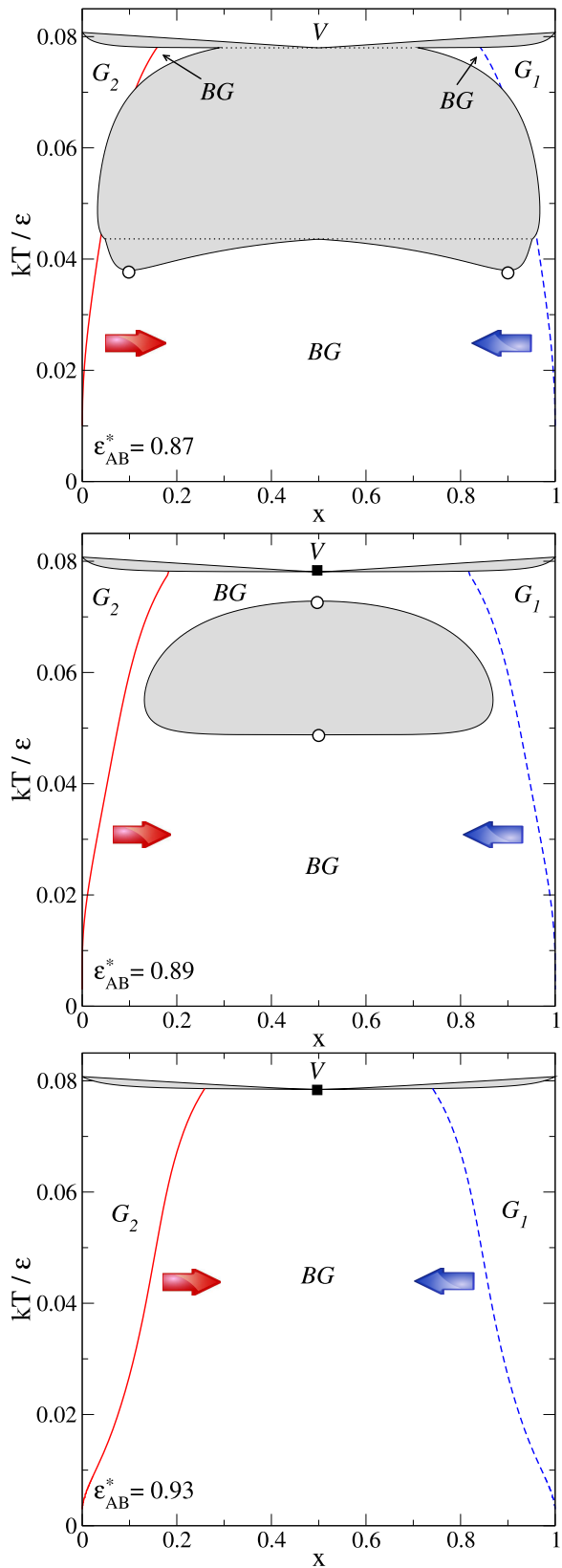,width=3.in,clip=}
\caption{Caption as in Fig. \ref{fig0.83} for mixtures with: $\varepsilon_{AB}^*=0.87$ (top), $\varepsilon_{AB}^*=0.89$ (middle), and $\varepsilon_{AB}^*=0.93$ (bottom).}
\label{figab8789}
\end{figure}

As expected, the demixing region decreases as the interaction between the $A$ and $B$ sites increases. At $\varepsilon_{AB}^*=0.87$ (see Fig. \ref{figab8789} top) the demixing 
region is still connected to the network fluid-vapor binodal. At intermediate temperatures there is a triple point where three bicontinuous gels at different compositions coexist. 
Below the triple point there are two symmetric and small regions of phase coexistence of two BGs, which end at lower critical points. At $\varepsilon_{AB}^*=0.89$ (see Fig. \ref{figab8789} middle) the demixing region is a closed loop of immiscibility bounded by upper and lower critical points. At $\varepsilon_{AB}^*=0.93$ (see Fig. \ref{figab8789} bottom) 
the demixing region has completely disappeared. There is no phase separation at temperatures below the network fluid-vapor binodal. 

Let us focus now on the percolation threshold(s). In all cases depicted in Fig. \ref{figab8789}, the percolation lines intercept the network fluid-vapor binodal at high temperatures 
and tend asymptotically to $x=0$ and $x=1$ as the temperature vanishes. It is possible to find a $BG$ at any composition if the temperature is sufficiently low. At compositions near $x=0$ and $x=1$ (pure fluids) the stable phase is a standard gel, $G_1$ or $G_2$, where only one species percolates. These regions grow as the interaction between the $A$ and $B$ 
sites increases. As a result, the intermediate region (where the $BG$ is stable) decreases. This is to be expected as the fraction of particles with more than one bond between 
different species increases as $\varepsilon_{AB}$ increases. 

Considering only the percolation thresholds, we conclude that the stability of the bicontinuous gel increases as the interaction between the $A$ and $B$ sites decreases. However, 
at low $\varepsilon_{AB}$ there is a two standard gels demixing region, which effectively reduces the stability of the $BG$ phase. Consequently, there is an optimal value of $\varepsilon_{AB}$ that maximizes the thermodynamic stability of the $BG$ structure. 

\subsubsection{$\varepsilon_{AB}\approx\varepsilon$}

\begin{figure}
\epsfig{file=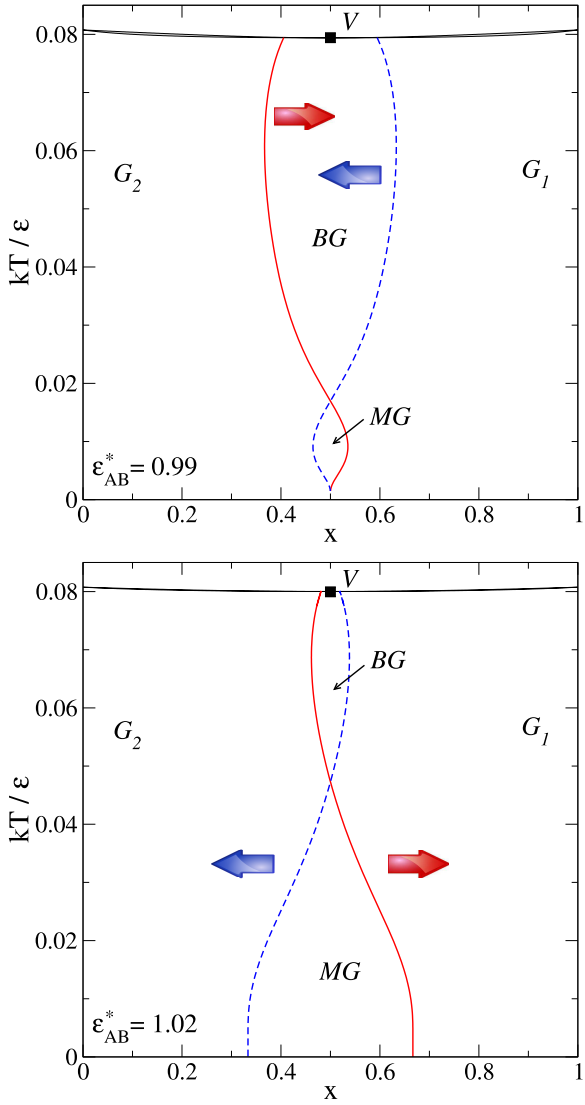,width=3.in,clip=}
\caption{Caption as in Fig. \ref{fig0.83} for mixtures with: $\varepsilon_{AB}^*=0.99$ (top), $\varepsilon_{AB}^*=1.02$ (bottom).}
\label{figab99102}
\end{figure}

The most interesting systems correspond to mixtures where the energy of the $AB$ bonds (between unlike species) is similar to the energy of the $AA$ and $BB$ bonds (between like species). In Fig. \ref{figab99102} we illustrate the phase diagram of a binary mixture with $\varepsilon_{AB}^*=0.99$ (top) and $\varepsilon_{AB}^*=1.02$ (bottom). In both cases we 
find the four different gel structures by varying the temperature and/or the composition of the mixture. In the pure fluid regime ($x\approx0$ and $x\approx1$) only one species is percolated giving rise to standard $G_1$ or $G_2$ gels, while in the equimolar regime around $x=0.5$ there is a competition between bicontinuous and mixed gels. The bigel is stable 
at high temperatures and the mixed gel at low temperatures. The stability of the $BG$ or $MG$ is again the result of a competition between the entropy and the energy of bonding.

Let us start by describing the mixture with $\varepsilon_{AB}^*=1.02$, that is, all bonds between different species ($AB$ and $CD$) are stronger than the bonds between particles of 
the same species ({\it e.g.} $AA$, $AC$ or $CC$). We expect a stable mixed gel (or standard gels if the composition is far from $x=0.5$). In fact, a $MG$ appears at low temperatures, but at high temperatures there is a small region where a $BG$ is stable. The stability of the $BG$ results from a gain in the entropy of bonding. Consider a particle of species $1$. 
It has $2$ sites of type $A$ which form $AA$, $AC$ and $AB$ bonds, and $1$ site of type $C$ which forms $CC$, $CA$ or $CD$ bonds. There are $27$ different combinations when all sites are bonded. $7$ of these favor the formation of mixed gels (those with $2$ or $3$ bonds between different species) while $20$ favour bicontinuous gels (those with no more than $1$ bond between different species). At high temperatures, the entropic part of the bonding free energy dominates and it is possible to stabilize bicontinuous gels even when the $AB$ bonds are stronger than, for example, $AA$ or $AC$ bonds. The opposite behaviour (a stable mixed gel when the $AB$ bonds are weaker than the $AA$ or $AC$ bonds) is also possible. In Fig. \ref{figab99102} (top), we illustrate the phase diagram of a mixture with $\varepsilon_{AB}^*=0.99$. At high temperatures the $BG$ is stable but at low temperatures we find a small region of stability of the $MG$. Finally, at zero temperature, the $BG$ reappears at $x=0.5$ (note that the ground state, at $x=0.5$, is a bicontinuous gel since there are only $CD$ and $AA$ or $BB$ bonds in order to minimize the energy). The $MG$ is stable in a small region at low temperatures because it minimizes the energy of bonding. Recall that there are $7$ combinations of bonds that favour a $MG$. When $\varepsilon_{AB}^*=0.99$ the average energy per bond of these configurations is $\langle\varepsilon^*_{MG}\rangle\approx1.007$, while the average over the configurations that favor a $BG$ is $\langle\varepsilon^*_{BG}\rangle=1.002$.  

\begin{figure}
\epsfig{file=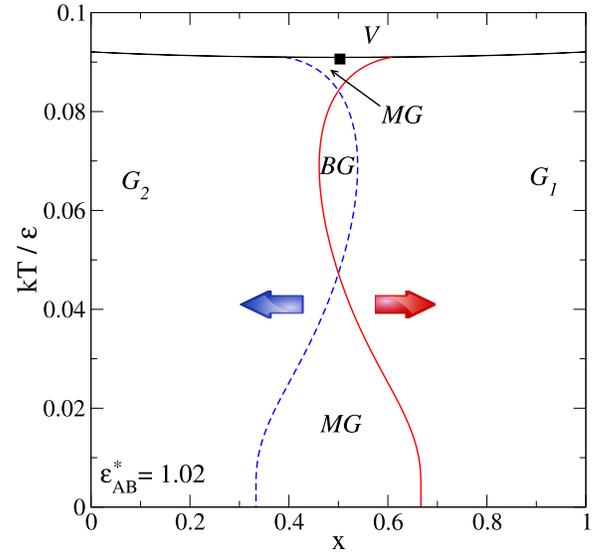,width=3.in,clip=}
\caption{Reduced temperature-composition phase diagram at constant pressure for an $AAC-BBD$ binary mixture with $\varepsilon_{AB}^*=1.02$. The pressure is $p^*=pv_s/\varepsilon=5.24\times10^{-4}$, slightly below the critical pressure of the pure fluids.}
\label{figab102p2}
\end{figure}

If the above argument is correct, we should find a stable $MG$ at low temperatures when $\varepsilon_{AB}^*=0.975$ (at this value of $\varepsilon_{AB}$ the mixed and bigel average energies per bond are equal, $\langle\varepsilon^*_{BG}\rangle=\langle\varepsilon^*_{MG}\rangle=1$). However, the $MG$ appears when $\varepsilon_{AB}^*\approx0.977$ (at the given pressure). This discrepancy may be traced to the assumption that all the sites are bonded, which is only an approximation at finite $T$. In fact, the fraction of unbonded sites can 
play a major role in the stability of percolated structures. For example, at high temperatures and pressures slightly below the critical pressure of the pure fluids, the fraction of unbonded sites is relatively high. Then, a significant fraction of particles will have only one or two sites bonded, increasing the number of configurations that favor a mixed gel. 
As a result the $MG$ becomes stable at temperatures above the region of stability of the $BG$. An example is shown in Fig. \ref{figab102p2}, where the interaction between sites $A$ 
and $B$ is $\varepsilon_{AB}^*=1.02$ (the same as in Fig. \ref{figab99102} (bottom)) at a pressure $p^*=pv_s/\varepsilon=5.24\times10^{-4}$ (slightly below the critical pressure of 
the pure fluids). Lowering the temperature at $x=0.5$ we find a reentrant sequence of gel structures: $MG-BG-MG$.

\begin{figure}
\epsfig{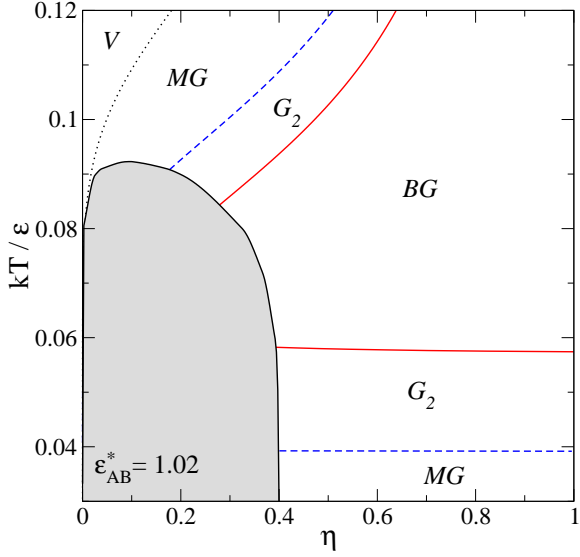}
\caption{Reduced temperature-packing fraction phase diagram at constant pressure of the symmetric $AAC-BBD$ binary mixture at composition $x=0.47$. The interaction between the $A$ 
and $B$ sites is $\varepsilon_{AB}^*=1.02$. A tie line connecting two coexisting phases is out of this plane.} 
\label{figxcte}
\end{figure}

In Fig \ref{figxcte} we represent a different phase diagram for the mixture with $\varepsilon_{AB}^*=1.02$. This is a cut of the $(T,\eta,x)$ phase diagram at constant composition $x=0.47$ (note that in this representation a tie line connecting two coexisting points is out of the plane). The dotted-black line is the percolation threshold of the mixture 
(not shown in the other phase diagrams since the fluid phase is always percolated).

\subsubsection{$\varepsilon_{AB}\gg\varepsilon$}

\begin{figure}
\epsfig{file=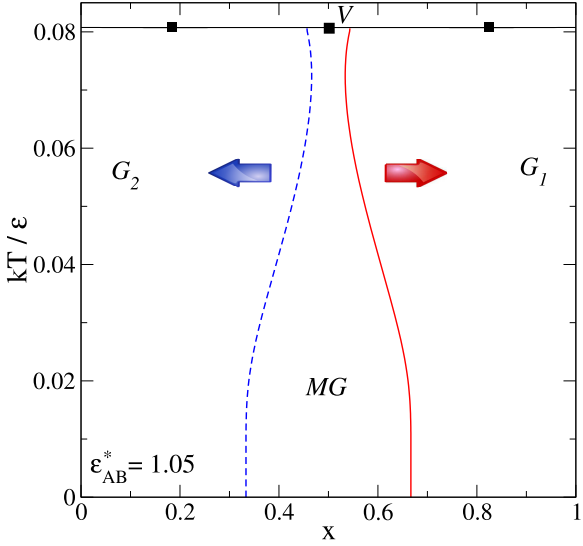,width=3.in,clip=}
\caption{Caption as in Fig. \ref{fig0.83} for mixtures with: $\varepsilon_{AB}^*=1.05$.}
\label{figab105}
\end{figure}

When the interaction between sites $A$ and $B$ is much stronger than the interaction between sites of the same species, we expect only mixed and standard gels. We finish this section with a simple case where $\varepsilon_{AB}^*=\varepsilon_{CD}^*=1.05$, that is, all the bonds between different species are stronger than the bonds between identical particles. This system is similar to the symmetric binary mixture $AAA-BBB$, analysed in section \ref{AAABBB}, but not identical. The phase diagram is depicted in Fig. \ref{figab105}. There are $3$ azeotropic points on the network fluid-vapor binodal ($2$ positive azeotropes at $x\approx0.18$ and $x\approx0.82$, and a negative azeotrope at $x=0.5$). The mixture is completely stable below the network fluid-vapor binodal temperature. A mixed gel is stable in a broad region around $x=0.5$. This region is bounded on the left by a $G_2$ structure and on the right by a $G_1$ structure. As expected, no $BG$ structure is found.

\subsection{Asymmetric binary mixture AAC-BBD}\label{AACBBDa}

\begin{figure}
\epsfig{file=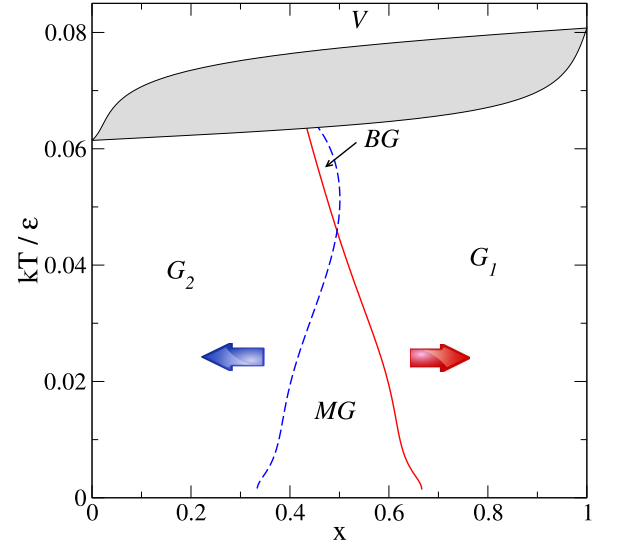,width=3.in,clip=}
\caption{Reduced temperature-composition phase diagram at constant pressure ($p^*=4.19\times10^{-5}$) of an asymmetric $AAC-BBD$ binary mixture. See the text for details 
of the bonding interactions.} 
\label{figasi}
\end{figure}

The phenomenology described above is not restricted to symmetric binary mixtures, which were considered for simplicity only. In Fig. \ref{figasi} we show an example of a similar 
phase diagram for an asymmetric binary mixture. It corresponds to a mixture where the symmetry is broken through the interactions between the sites on particles of species $2$. The 
set of interaction energies is: $\varepsilon_{AA}=\varepsilon_{AC}=\varepsilon_{CC}=\varepsilon$ (it sets the scale of energy), $\varepsilon_{BB}=\varepsilon_{BD}=\varepsilon_{DD}=0.75\varepsilon$, $\varepsilon_{CD}=0.95\varepsilon$, $\varepsilon_{AB}=0.88\varepsilon$ and $\varepsilon_{AD}=\varepsilon_{BC}=0$. The mixture is always percolated at temperatures below the network fluid-vapor binodal (which now occupies a significant region of the phase diagram). By varying the temperature and composition of the mixture it is possible to find the four percolated structures described for symmetric mixtures. As before, the stability of the different structures can be 
understood in terms of the competition between the energy and the entropy of bonding.

\section{Conclusions}\label{Conclusions}

We have studied the thermodynamics and percolation thresholds of a simple model of patchy colloidal particles: a binary mixture of hard spheres with different types of bonding 
sites on their surfaces. Despite the simplicity of the model, the mixture exhibits a very rich phase diagram. The network fluid regime includes up to four gel phases, including bicontinuous and mixed gels. The bicontinuous gels described here are similar to those studied by Hall an co-workers \cite{B907873H,goyal:064511} in mixtures of dipolar colloids.

The stability of the different percolated structures is determined by a competition between the entropy of bonding (number and type of bonding sites) and the energy of bonding (bonding energy). Therefore, we expect the results to be relevant to a wide range of patchy particle systems, regardless of the size and/or the geometry of the colloids. 

We have shown that by tuning the bonding energies it is possible to design binary mixtures with stable bigels and/or mixed gels in a wide range of temperatures and 
compositions. When the strength of the bonds between like and unlike particles is similar, we find an interesting competition between the mixed and bicontinuous gels, including 
reentrant behaviour. The mixtures where both bicontinuous and mixed gels compete are promising candidates to fabricate materials with novel physical properties. An obvious example 
is a mixture where one of the species transmits a given property (light, electricity...) while the other does not. It is possible to control the transmission of that property by 
simply varying the temperature, in the regime where $MG$ and $BG$ compete.

We have not considered here the stability of positionally ordered phases. Solid phases will appear at low temperatures and/or high pressures preempting part of the phase diagrams. Nevertheless, the gels we have analysed appear at relatively low packing fractions. In addition, it is expected that the random distribution of patches on the particle surfaces 
will frustrate the formation of solid structures.
 
\section{Acknowledgments}

This work has been supported, in part, by the Portuguese Foundation for Science and Technology (FCT) through Contracts Nos. POCTI/ISFL/2/618 and PTDC/FIS/098254/2008, by the 
R$\&$D Programme of Activities (Comunidad de Madrid, Spain) MODELICO-CM/S2009ESP-1691, and by the Spanish Ministry of Education through grant FIS2010-22047-C05-01. D. de las Heras 
is supported by the Spanish Ministry of Education through contract No. EX2009-0121.

\end{document}